%% file: main.tex
\documentclass[conference]{IEEEtrans}
\IEEEoverridecommandlockouts
\usepackage{cite}
\usepackage{xcolor}
\usepackage{amsmath,amssymb,amsfonts}
\usepackage{algorithmic}
\usepackage{graphicx}
\usepackage{textcomp}

\usepackage{hyperref}
\usepackage{cleveref}
\usepackage{prettyref}
\usepackage{comment}
\usepackage{dblfloatfix}
\usepackage{multirow}
\usepackage{placeins}
\usepackage{harveyballs} 
\usepackage{paralist}
\usepackage{xspace}

\usepackage[sort&compress,square,numbers]{natbib}

\usepackage{lipsum}

\usepackage{todonotes} 
\newcommand{\xynote}[2]{\todo[inline]{#1: #2}}

\newcommand{\jz}[1]{\xynote{Jingxi}{#1}}

\newcommand{\aas}[0]{Asset Administration Shell\xspace}

\def\BibTeX{{\rm B\kern-.05em{\sc i\kern-.025em b}\kern-.08em
    T\kern-.1667em\lower.7ex\hbox{E}\kern-.125emX}}
\begin{document}

\title{A Container-based Approach For Proactive Asset Administration Shell Digital Twins}

\author{\IEEEauthorblockN{Carsten Ellwein, Jingxi Zhang, Andreas Wortmann}
\IEEEauthorblockA{\textit{Institute for Control Engineering of Machine Tools} \\
\textit{and Manufacturing Units (ISW)}\\
\textit{University of Stuttgart}\\
Stuttgart, Germany \\
{\{carsten.ellwein,  jingxi.zhang, andreas.wortmann\}}@isw.uni-stuttgart.de}
\and
\IEEEauthorblockN{Antony Ayman Alfy Meckhael}
\IEEEauthorblockA{{\textit{Faculty of Media Engineering }} \\ \textit{and Technology (MET)}\\
\textit{German University in Cairo}\\
New Cairo City, Egypt\\
{antony.alfy@student.guc.edu.eg}}
\thanks{Partly funded by the Federal Ministry for Economic Affairs and Energy (BMWE) through the projects growING (grant no. 13IPC036G).
Partly funded by the Deutsche Forschungsgemeinschaft (DFG, German Research Foundation) -- Model-Based DevOps -- 505496753.
The authors are responsible for the content of this publication.}
}

\maketitle
\input{00_Abstract}
\input{01_Introduction}
\input{02_RelatedWork}

\input{03_Architecture}
\input{04_CaseStudy}

\input{05_Conclusion}

\bibliographystyle{IEEEtrans}
\bibliography{bibliography}

\end{document}

%% file: 00_Abstract.tex
\begin{abstract}  
In manufacturing, digital twins, realized as Asset Administration Shells (AAS), have emerged as a prevalent practice. These digital replicas, often utilized as structured repositories of asset-related data, facilitate interoperability across diverse systems.
However, extant approaches treat the AAS as a static information model, lacking support for dynamic service integration and system adaptation.
The existing body of literature has not yet thoroughly explored the potential for integrating executable behavior, particularly in the form of containerized services, into or from the AAS. This integration could serve to enable proactive functionality.
In this paper, we propose a submodel-based architecture that introduces a structured service notion to the AAS, enabling services to dynamically interact with and adapt AAS instances at runtime.
This concept is implemented through the extension of a submodel with behavioral definitions, resulting in a modular event-driven architecture capable of deploying containerized services based on embedded trigger conditions. The approach is illustrated through a case study on a 3-axis milling machine.
Our contribution enables the AAS to serve not only as a passive digital representation but also as an active interface for executing added-value services.

\end{abstract}


%% file: 01_Introduction.tex
\section{Introduction}
\label{sec:introduction}

Digital twins~(DTs)~\cite{kritzinger2018digital,tao2018digital} are software systems~\cite{munoz2023conceptual} that connect to a cyber-physical system~(CPS), automatically receive data from it, perform computations and return instructions to the CPS ~\cite{kritzinger2018digital}. 
DTs include the physical entities (i.e. CPS), their data, models, and active software components~\cite{tao2018digital}.
An important modeling technology in the manufacturing environment is the Asset Administration Shell~(AAS), utilized modeling products~\cite{ajdinovic2024}, processes~\cite{dietrich2024}, resources~\cite{frick2024} and their interaction in the context of production~\cite{ellwein2023}.

The AAS models exist in an ecosystem; they can be provided on a server and refer to current values of the CPS in their structural description.
Therefore, according to the literature~\cite{wagner2017role, redeker2021towards, neubauer2023}, AAS (or rather the AAS ecosystem) could be used as a technology to implement DTs~\cite{zhang2025}.
However, as outlined in the related work section, the AAS ecosystem lacks a way to integrate active software components, thus proactive behavior, in an encapsulated and thus reusable manner.
Consequently, the lack of encapsulation and re-usability prevents the establishment of standardized or shared proactive AAS components.
This research gap is addressed in this paper.
The integration of a container runtime into the AAS ecosystem enables the encapsulated deployment of software components.
The concept presented in the body of the paper enables the container runtime behavior to be influenced via AAS, thus synchronizing software service, data and model.
Our contribution enables the AAS to serve not only as a passive digital representation but also as an active interface for executing added-value services.

The remainder of this paper is structured as follows:
Section~2 provides a comprehensive review of the extant literature on the AAS and the five-dimensional-digital-twin model. In Section~3, the architecture of the proposed solution is presented. In Section~4, the feasibility and benefits of this approach are demonstrated through a case study involving a 3-axis milling machine. Concluding the paper is Section~5, which outlines directions for future research.

%% file: 02_RelatedWork.tex
\section{Related Work}
\label{sec:RelatedWork}


%
%
The 5D-model~\cite{tao2018digital} delineates that DTs comprise five distinct elements: (i)~physical entity, (ii)~virtual models, (iii)~services, (iv)~data and (v)~the connections, which facilitates collaboration between the four previous components.
The (i) existence of physical entities within the physical world serves as the foundational basis for DTs.
These physical entities are governed by physical laws. 
DTs are designed to replicate the attributes of their corresponding physical entities, thereby enabling the simulation of their behaviors.
The utilization of (ii) virtual models has been proven to be an effective method of replicating both physical entities and the physical properties and behaviors that are inherent to these entities.
This includes behavior models describing how entities respond to changes in their environment, rule models providing logical abilities (e.g., reasoning or evaluation).
The (iii) services offered by DTs include a range of applications such as simulation, verification, monitoring, optimization, diagnosis, prediction, prognosis, and health management.
The (iv) data is at the core of the model and originates from all the previously mentioned elements.
As the (v) connections, data can be created persistently, but can also be fleeting.

%
%
One technology that can be used to implement DTs is the AAS~\cite{zhang2025}.
The AAS is defined as the digital representation of the asset containing all its relevant information throughout its entire lifecycle~\cite{DinSpec.91345} and is presented as the basis of interoperability: on the one hand, it holds information of various types and on the other hand, it functions as the interface for communication within the I4.0 network through which information can be exchanged between assets~\cite{ZVEI.2019}.  

Since the AAS holds relevant information throughout the lifecycle of the asset, it must be capable of representing different sorts of information, such as properties, modeled functionalities, parameters, a summary of included components, as well as data that accrues during manufacturing or simulation and also descriptions, such as their usage instructions and technical specifications.  
This presupposes the ability to store or refer to heterogeneous data and models~\cite{cavalieri2020asset}. In relation to the 5D-model, these represent the virtual model. 
The AAS is structured within a metamodel which defines structure and semantics of an AAS, such as the way elements are related or identified. This metamodel is instantiated in specific AAS submodels.
Each individual submodel of an AAS is intended to represent one content-related or functional aspect of the represented asset.
Submodels can be created individually applying the previously introduced metamodel. 
To ensure consistency and interoperability, the Industrial Digital Twin Association~(IDTA) provides standardized and publicly available so-called submodel templates, which we extend in our concept to enable proactive services in AAS, thus showing an approach for services in DTs. 
There is an increasing focus on the representation of software components in digital twin architectures, a trend that is also evident in recent advancements within the IDTA submodel ecosystem. 
For instance, submodels such as the Software Nameplate and the Software Bill of Material primarily support documentation and traceability during runtime. 
Conversely, submodels such as Interface Connectors and Computing Platform Resources furnish structural and deployment-related information that can be leveraged during the preliminary engineering phases, including system integration and configuration.
Parallel to the stronger focus on the description of software components, the software-heavy nature of the AAS itself is also increasing. 
The declared aim of some of the projects described in more detail below is to provide AASs with active behavior, the so-called proactive asset administration shells.

%
%
One work~\cite{grunau2022implementation} proposes an approach to extend the inherently reactive behavior of the AAS by integrating a proactive component. 
Their solution involves engineering a tightly coupled software module that operates in conjunction with an AAS server. The demonstration shows two AAS, a product and a storage facility, and multiple bidding apps, that calculate prices for production and storage, monitor the AAS and select appropriate actions. 
Although effective in demonstrating basic proactivity, the approach entangles software engineering logic with domain-specific knowledge, resulting in limited modularity and a high degree of manual effort on service implementation. This tight coupling complicates reuse and scalability, especially when adapting services in different asset types or operational contexts. 

One work~\cite{longo2019ubiquitous} proposes a server-based middleware architecture in which a centralized service bus mediates interactions between services via RESTful web services. Although this approach provides a network accessible interface to digital twin functionality, the paper lacks a detailed description of internal execution flows or lifecycle management of services within the middleware. Additionally, although the authors' ontology-based knowledge structure resembles the role of multiple submodels in an AAS, their implementation lacks modularization; all software components are tightly coupled with a single middleware. 

Another relevant contribution is to AI-powered digital twins~\cite{bolbotinovic2025ai}, which explore the use of runtime-enriched digital twins for anomaly detection. Their approach uses live operational data to train and apply AI models within a twin framework, that they propose.
However, their solution does not align with AAS standards, resulting in a monolithic system architecture with tightly integrated, non-modular methods. 
Consequently, the proposed approach requires more effort to generalize the approach beyond their case study. This highlights a key limitation in current AI-integrated twin research: the absence of standardized, interface-oriented structures that enable service composition and reuse. Our work addresses this gap by introducing an AAS-compliant, submodel-based mechanism for embedding containerized, AI-ready services into digital twin systems.

\begin{figure}
    \centering
    \includegraphics[width=0.49\linewidth]{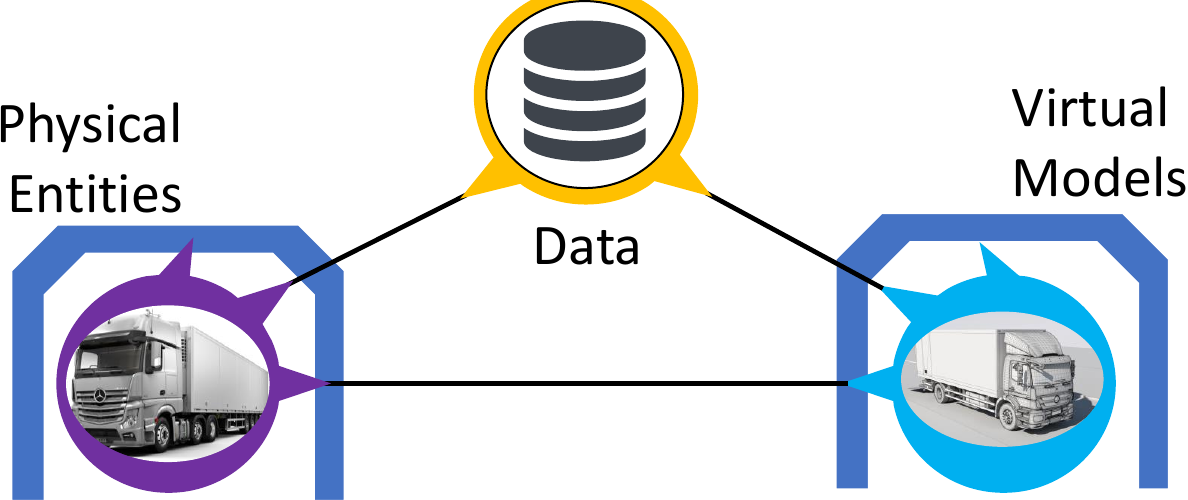}
    \includegraphics[width=0.49\linewidth]{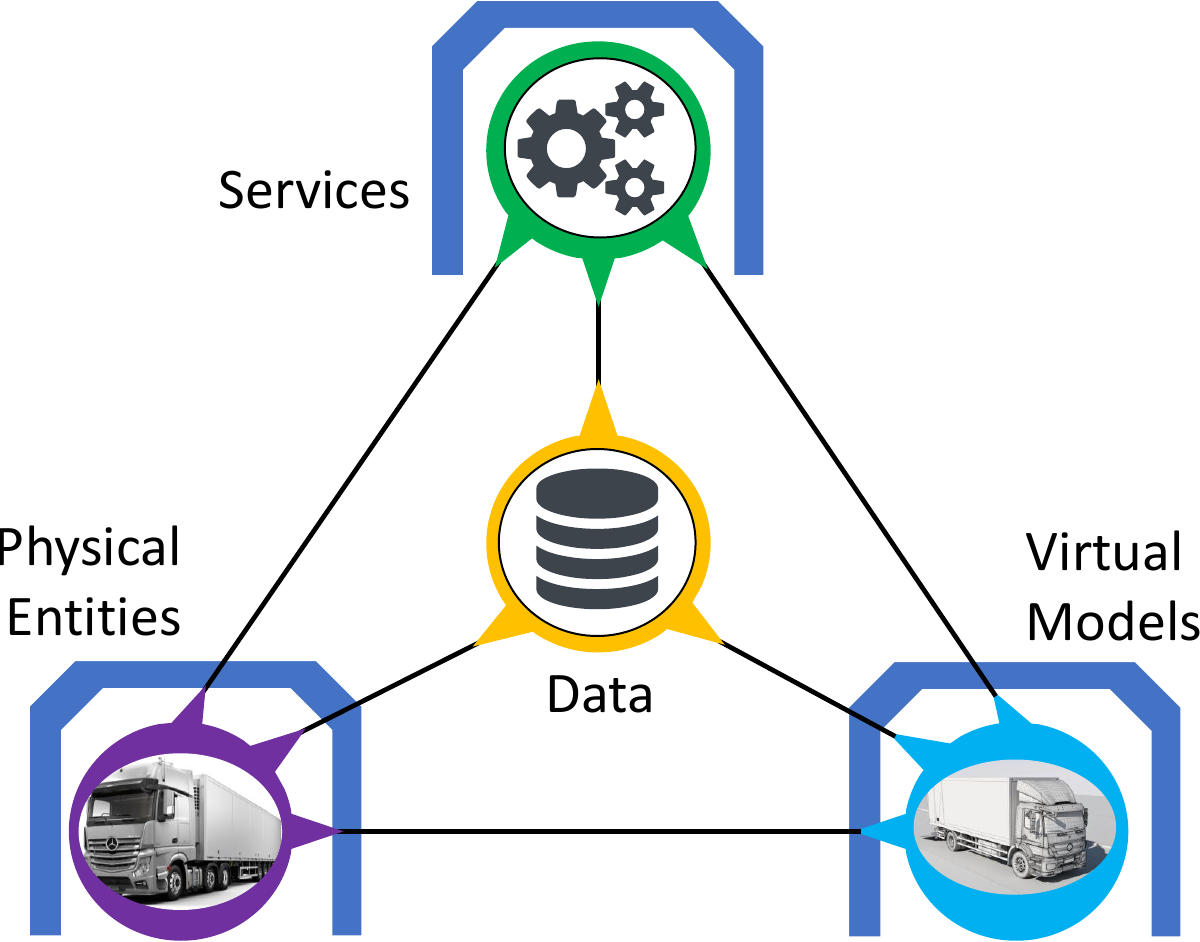}
    \caption{The current state of the community effort for the development of AAS \textit{(left)} vs. the aspired goal within this work \textit{(right)}, based on \cite{tao2018digital}.}
    \label{fig:AAST_T3}
\end{figure}

%
%
In summary, existing approaches to the AAS predominantly treat it as a static information model, lacking active components capable of adapting or influencing the system at runtime. 
Integrated into the 5D-model, current AAS implementations cover the physical entity and their models~\cite{zhang2025} (cf. Fig.~\ref{fig:AAST_T3}~\textit{(left)}).
The absence of a structured notion of services within the AAS limits its ability to support proactive behavior or dynamic system adaptation. 

As software becomes increasingly complex, but also central to industrial systems, this imbalance introduces a critical need for models and interaction patterns that are not tied to concrete machine instances. 

To address this, we propose an extension of the AAS that integrates executable services as first-class elements (cf. Fig.~\ref{fig:AAST_T3}~\textit{(right)}).
This enables both human users and automated algorithms to interact with and adapt the system through the AAS interface. 
By establishing a foundation for proactive behavior, our approach opens the door to future developments in AI-driven decision-making and added-value services within digital twin ecosystems.

%% file: 03_Architecture.tex
\section{Architecture Layout}
\label{sec:Architecture}
In order to equip the AAS with proactive software capabilities while staying in conformity with the AAS standard~\cite{idtaAASSpecPart2}, which requires an explicit definition of identifiers, relations, versioning, and submodels, it is first necessary to identify suitable submodels or component representations that can serve as integration points for added-value services. 

In order to provide support for the dynamic deployment of containerized services based on the AAS standard, a modular, event-driven architecture is further introduced. This architecture enables runtime orchestration and contextual service activation. 

The fundamental rationale behind this design choice pertains to the objective of decoupling the standardized management of AAS submodels from the specialized logic that is responsible for orchestrating the lifecycle of external containerized services. This decoupling facilitates system extensibility, maintainability, and modular integration of dynamic service behavior~\cite{stutterheim2018maintaining}.

\begin{figure}
    \centering
    \includegraphics[width=\linewidth]{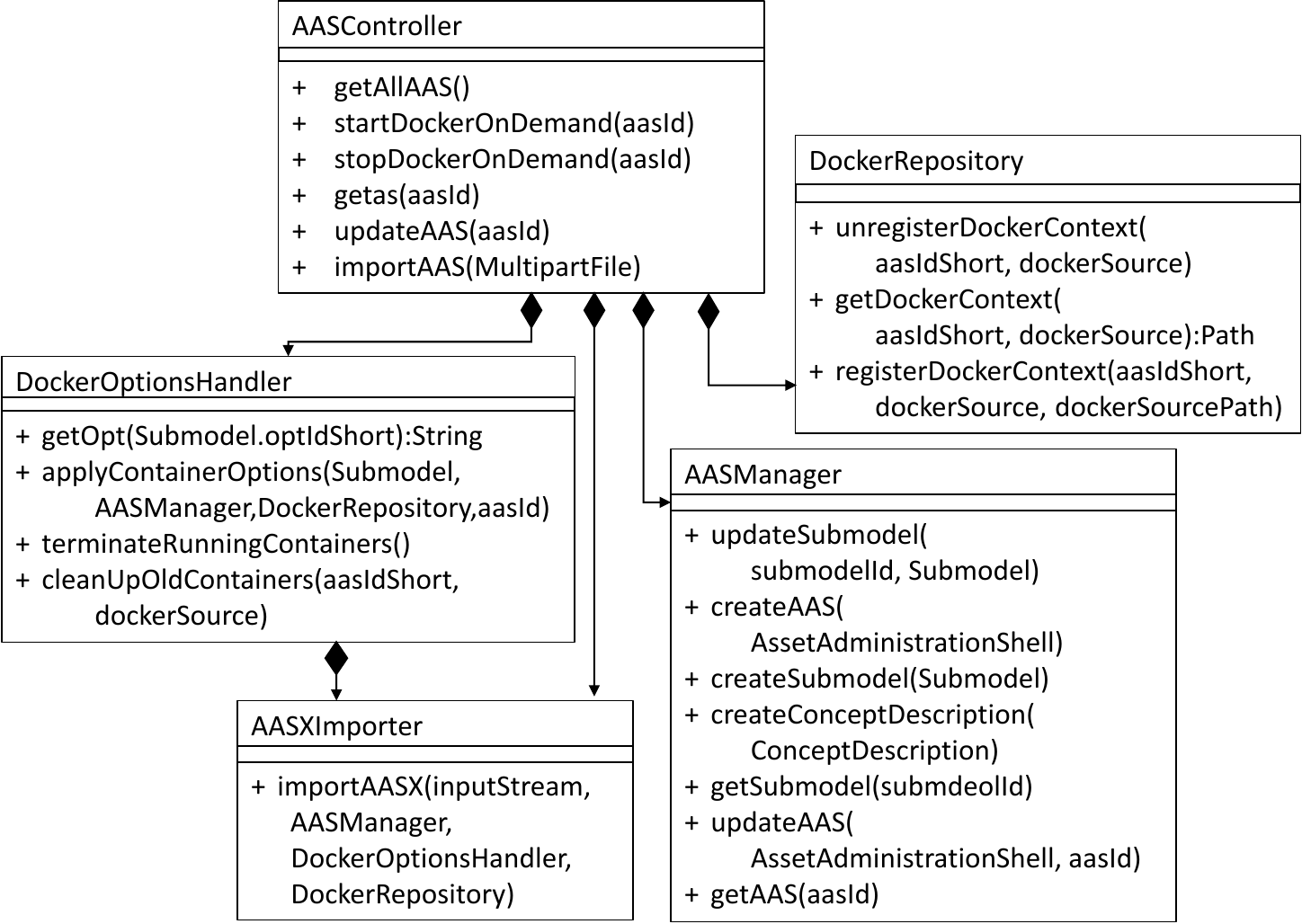}
    \caption{Conceptual component diagram of the architecture, illustrating the primary components and their logical dependencies.}
    \label{fig:class_diagram}
\end{figure}

Our system is designed to operate on the so called AASX package, a ZIP-based archive intended for the structured exchange of AAS submodels along with associated supplementary files.
A required part of this archive is an XML-based representation of the supplementary files contained and information on the submodel.
We extend this convention by introducing a method for co-locating behavioral definitions alongside declarative submodel elements within the same package.
Such an extension consists of the versioning and execution of the service, which can be exported to different AAS. 
\paragraph{The Service Execution Submodel}
To enable the discovery and interpretation of service execution logic, a dedicated submodel, the Service Execution Submodel, is introduced. It provides a standardized structure for specifying container-related metadata, such as execution triggers and termination policies. The formal specification, a template, and an example AASX package for this submodel are publicly available \footnote{Service Execution Submodel available via \url{https://doi.org/10.18419/DARUS-5304  }.}.
In this sense the service execution submodel represents added-value services, capable of operating on existing AAS instances. Consequently, it is imperative that the system under consideration provides support for explicit references to both internal and external AAS, their constituent submodels, and the specific submodel elements that are indispensable for the system's operation. Furthermore, a description of the service's behavior is needed. To this end, we propose the representation of such behavior as executable software artifacts, technically realized as containerized packages (e.g., Docker images), thereby enabling consistent deployment and integration.
\paragraph{Architecture of the system}
The architecture leverages the general concept of using containers to encapsulate service logic. For the specific implementation and use case presented in this paper, we utilize Docker as the container technology.
As such the core architectural components responsible for processing these enriched AASX packages are depicted in Fig.~\ref{fig:class_diagram}, which contains the components AASX Importer, AAS Manager, Docker Repository, Docker Options Handler, and AAS Controller.

The responsibilities of the core logical components in the architecture are defined as follows:
\begin{itemize}
\item \textbf{AAS Controller}: The controller functions as the principal interface through which users or services can access or manipulate submodels within an AAS. It also serves as the invocation interface for containerized services, which have the capacity to dynamically update or modify submodel elements during their execution.

\item \textbf{AASX Importer}: The primary responsibility of the importer is the processing of incoming AASX packages. This component facilitates the co-location of models and their corresponding service within the same AAS as well as across different AASs. Our approach employs a convention that involves the storage of contexts within a predefined directory structure.
The importer then deserializes the AAS submodels and routes both the model data and executable artifacts to their respective processing components.

\item \textbf{AAS Manager}: This component functions as the primary interface for the management of the lifecycle of AAS submodels, offering Create, Retrieve, Update, and Delete operations. The system's functionality encompasses the abstraction of the persistence layer, thereby ensuring the uniform accessibility of core AAS resources.

\item \textbf{Docker Repository}:  This repository is intended for the management of containerizable artifacts, such as the services behavior as Dockerfiles and the associated build contexts. The repository facilitates versioned retrieval and reuse of service logic defined within AASX packages.

\item \textbf{Docker Options Handler}: This handler functions as the runtime execution engine. It interprets service execution metadata embedded in AAS submodels and manages the full lifecycle of corresponding containers, from build to deployment and termination.
\end{itemize}

In the context of lifecycle management of services, with particular emphasis on containerized services such as Docker containers, it is imperative to delineate the temporal sequence of pivotal phases, encompassing build, deployment, and termination. To address this issue, our concept introduces an event-driven workflow specification, thereby enabling precise control over the execution lifecycle of services. 
A logical execution sequence is provided to determine the point in time at which the container is invoked, created, or terminated.

\begin{figure}
    \centering
    \includegraphics[width=\linewidth]{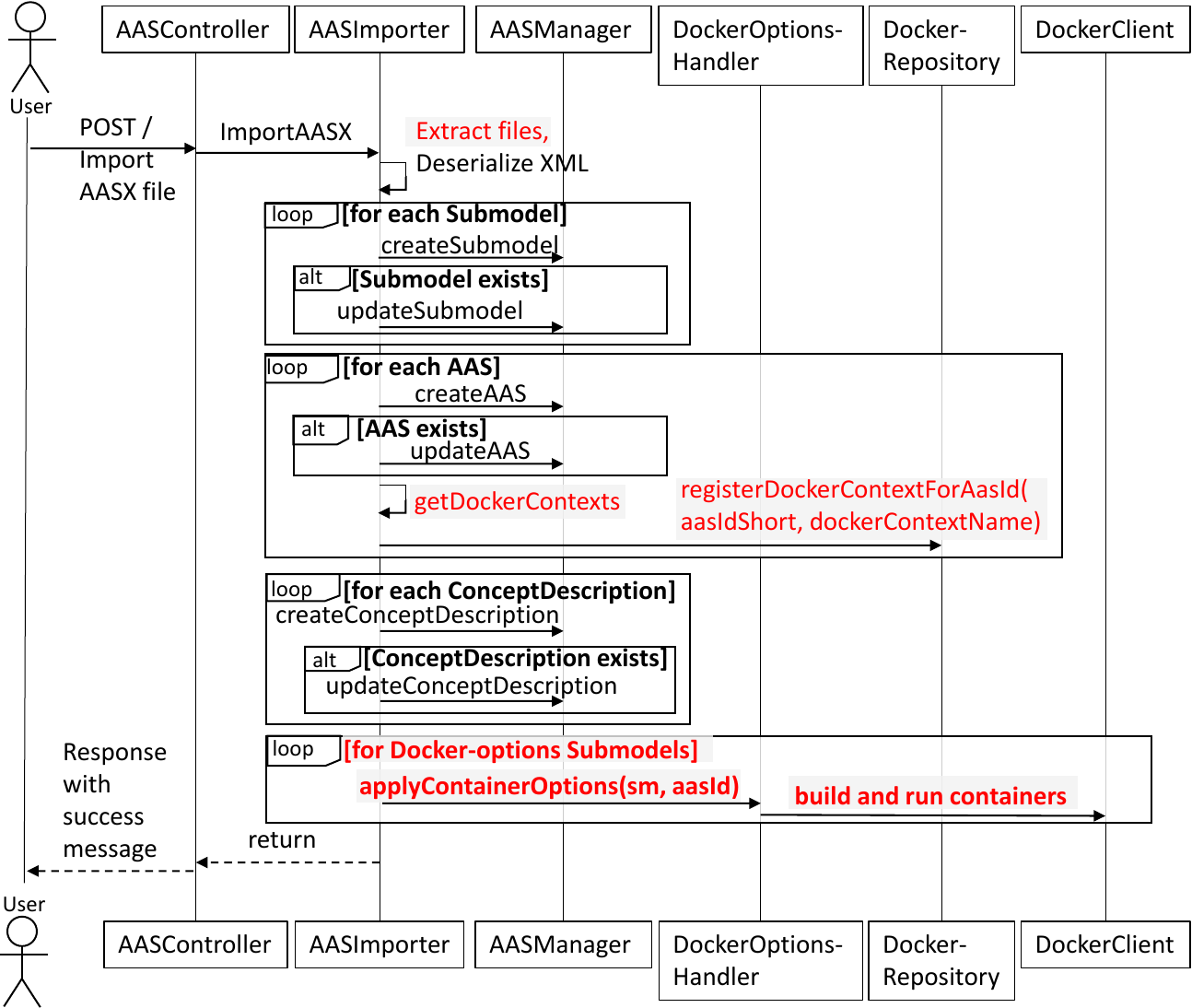}
    \caption{Sequence diagram of the import and service activation workflow.}
    \label{fig:sequence_diagram}
\end{figure}

\paragraph{Submodel based Execution Flow}
The operational workflow is initiated by the import of an AASX package into the AASX Importer, which subsequently instigates the registration and deserialization of its contents. This approach aligns with prevailing standards in the field. The present addition to the aforementioned framework pertains to the checking of service contexts, and, when applicable, the activation, build, and run of a corresponding containerized service. This workflow as depicted in Fig.~\ref{fig:sequence_diagram} illustrates the manner in which architectural components interact to transform a static, declarative package into a runtime-executable service. 

The workflow proceeds logically as follows:
\begin{enumerate}
    \item An AASX package is imported and processed by the \textbf{\aas Importer}.
    \item The importer directs the \textbf{\aas Manager} to register all submodels. For all submodels that contain a service context, the importer now instructs the \textbf{Docker Repository} to store the service contexts.
    \item Finally, the importer provides any submodels containing service metadata, such as submodel references and the \aas ID to the \textbf{Docker Options Handler}, which evaluates the provided service execution triggers, defined in the service execution submodel, and, if conditions are met, interacts with a container engine to build and run the service.
\end{enumerate}
The triggers depend on the context that is given in its service execution submodel. In the following we investigate which kinds of triggers are required in the lifecycle of a service in the AAS.

\paragraph{Conditional Logic for Service Activation}

\begin{figure*}
    \centering
    \includegraphics[width=0.85\textwidth]{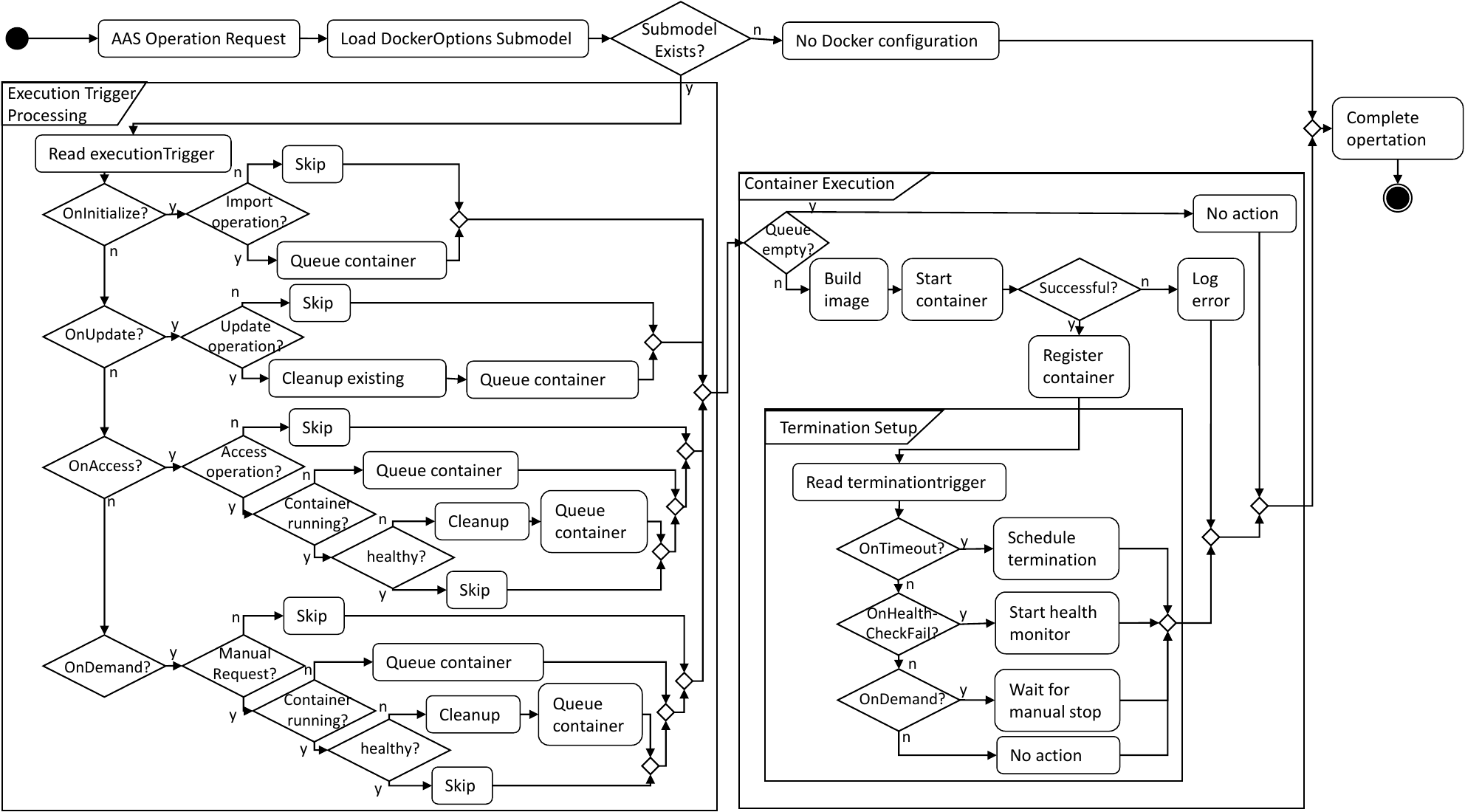}
    \caption{Activity diagram of the conditional logic for processing execution and termination triggers.}
    \label{fig:activity_diagram}
\end{figure*}

A fundamental aspect of the proposed architecture is its capability to respond to a predetermined set of triggers that are integrated directly within the AAS submodel. These triggers function as semantic activation points for containerized services, thereby facilitating context-aware behavior. The conditional logic that governs the evaluation and handling of these triggers is modeled in the activity diagram presented in Fig.~\ref{fig:activity_diagram}. 

The system evaluates \texttt{executionTrigger} properties based on the context of the operation, providing usage guidelines for different scenarios.
The following selection of execution triggers are supported in the current \texttt{ServiceExecutionSubmodel}:
\begin{itemize}
    \item \texttt{onInitialize}: Used for one-time setup tasks that should run when an asset is first introduced to the system (during initial import).
    \item \texttt{onUpdate}: Designed for services that must react to changes in the AAS data. This trigger ensures that a service is re-executed to maintain data consistency or re-evaluate a state whenever the source AAS is modified.
    \item \texttt{onAccess}: Suited for resource-intensive or just-in-time services that should only be activated when a client actively interacts with the asset, thereby conserving system resources until they are explicitly needed.
    \item \texttt{onDemand}: Reserved for services that require explicit external invocation through the API and shouldn't be triggered automatically by other lifecycle events.
\end{itemize}
This model-driven, declarative approach allows the behavior of an asset to be specified entirely within its digital twin, enabling a more autonomous and integrated cyber-physical system. 
Furthermore, by processing \texttt{terminationTrigger} metadata, the system provides complete lifecycle management, enforcing policies such as automated termination after a timeout (\texttt{onTimeout}) or in response to a failed health check (\texttt{onHealthCheckFail}).

In summary, our proposed service execution submodel, architecture, and triggers complement the advancements towards proactive AAS. 

%% file: 04_CaseStudy.tex
\section{Case Study}
\label{sec:CaseStudy}
In our case study we investigate the programming of CNC milling machines, which frequently necessitates a considerable amount of manual effort to adapt and calibrate the control logic for each individual machine instance. It is imperative to note that machines of the same type and configuration (e.g., multiple 3-axis or 4-axis milling machines) generally necessitate distinct tuning and adjustment of control programs. This absence of transferability is predominantly attributable to geometric error and the subsequent necessity for machine-specific error compensation. It is widely acknowledged that geometric inaccuracy is a primary source of deviation in CNC machining. A geometric error compensation is a standard requirement in both research and industrial applications across machines with three or more axes.
As demonstrated by extant research~\cite{raksiri2004geometric}, geometric and cutting-force-induced errors vary considerably among individual machine instances, even among models that appear identical. Consequently, compensation strategies, such as the use of laser interferometry for error measurement and the application of neural network-based correction models, are typically tailored to a single machine. While these methods have been demonstrated to achieve high levels of accuracy for the specific machine on which they are developed, the resulting CNC programs and compensation algorithms are not directly transferable to other machines, even of the same type and configuration. This instance-specific behavior engenders a fundamental limitation in current practice: each machine necessitates individual calibration and model adaptation, impeding scalability and reuse in production environments.
\begin{figure}
    \centering
    \includegraphics[width=0.85\linewidth]{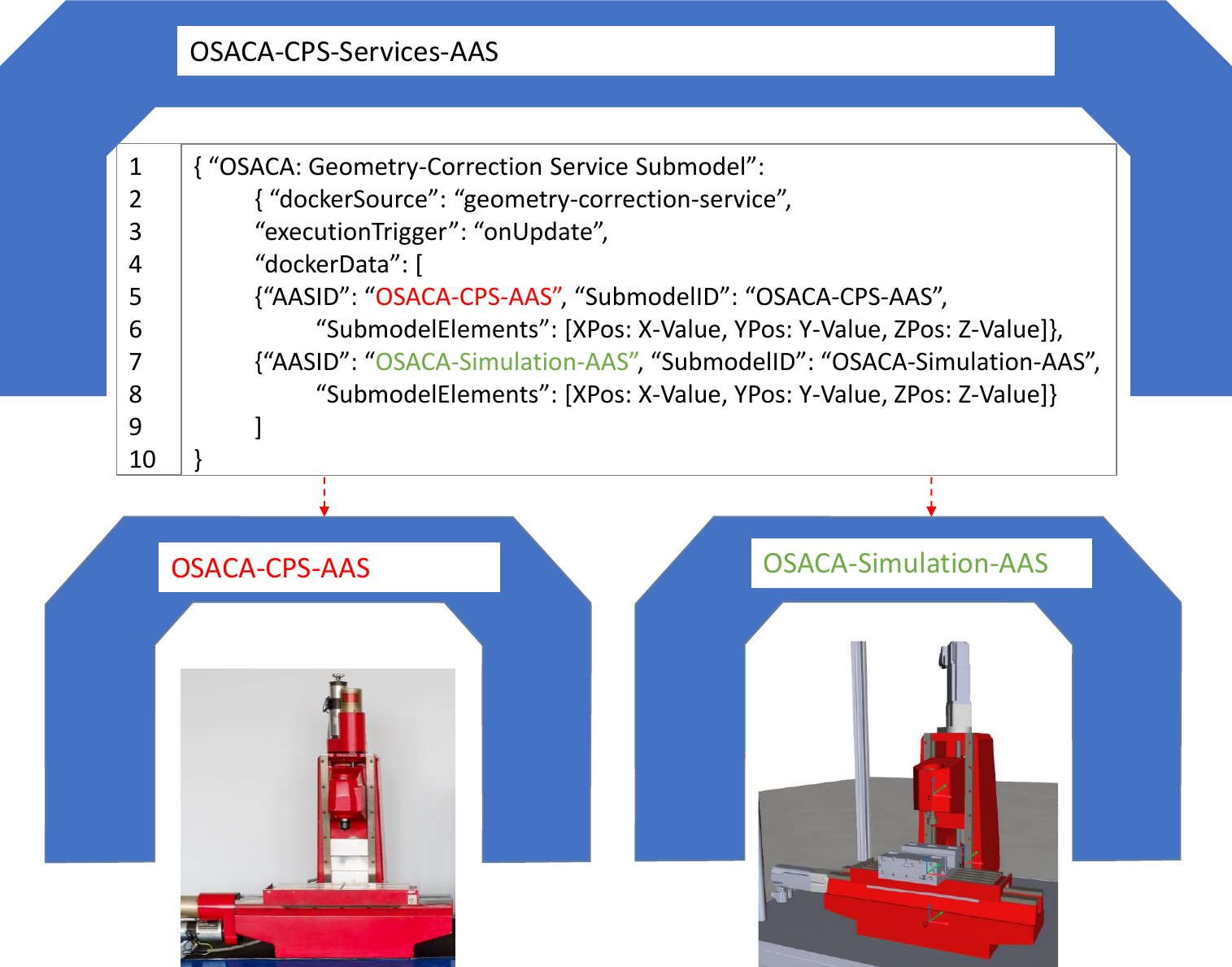}
    \caption{Constellation of three AAS: bottom left AAS of the physical machine containing the real values, bottom right AAS of the simulation model of the OSACA milling machine. On top a service that takes the values from the physical machine, uses a programmatic correlation in a Docker container and writes the normed value to the model AAS.}
    \label{fig:OSACA-AAS}
\end{figure}
As such, in our case study, we focus on a 3-axis CNC milling machine (cf. Fig.~\ref{fig:OSACA-AAS})~\cite{zhang2025} and use a simplified correlation model between the actual position and the modeled position. 

We will omit the factors of the laser measurement and the correlating deviation from the measurements. We assume a submodel which contains the observed value.
With the model of the positional error we develop a service to perform the compensation from the model to the real control.

This transformation algorithm is built in a dockerized service, which is invoked by the execution trigger \texttt{onUpdate}. Here an update may be a movement of the physical machine or an axis movement done to one of the axes which needs to be transferred to the physical machine.

To address the challenge of non-transferable and non reusable services due to machine-specific deviations, we use our service execution submodel to decouple machine-specific properties from generic control models. This objective is realized through the implementation of a service to establish a connection between the behavior of the physical machine and an abstract, idealized model representation. The architecture is implemented using three AAS in submodels depicted in Fig.~\ref{fig:OSACA-AAS}, each reflecting a distinct concern:
\begin{enumerate}
 
   \item The first submodel: This representation signifies the physical machine's current state in real time, which includes aspects such as the tool's position. It is intended for utilization by both machine experts and operators.

   \item Transformation submodel: The compensation logic is encapsulated as a service that corrects for geometric and process-induced errors.

   \item The model submodel is as follows: This model is an abstracted, machine-type-level representation that contains compensated values suitable for simulation, planning, or algorithmic reuse.

\end{enumerate}
This structured distinction of submodels supports software solutions that operate on different levels of abstraction, while remaining anchored to the instance level when needed. By abstracting compensated machine behavior, the submodel can be accessed by added-value services (e.g., predictive maintenance, adaptive path planning) that are no longer tightly coupled to a single machine instance.

The service we implemented is the geometry correction service, that follows the service execution submodel. The transformation service functions as a mediation layer, provided by a domain expert. The system performs a calculation to determine a compensated tool position for milling operations by accessing two AAS submodels. The OSACA-CPS-AAS submodel provides the tool's current measured position in reality. The OSACA-Simulation-AAS submodel, which defines the ideal or simulated position, is updated using the transformation logic.

Finally, the three submodels are deployed on an AAS server, which is implemented within a Java environment under the BaSyx framework. The server architecture comprises the following conceptual components: AAS Controller, AASX Importer, AAS Manager, Docker Repository and Docker Options Handler. This infrastructure provides full support for create, update and delete operations on AAS instances and submodels. To demonstrate the feasibility of this approach, we implemented an \texttt{onUpdate} trigger mechanism to enable consistent and automated propagation of position updates from the OSACA-CPS-AAS to the OSACA-Simulation-AAS.

This enables the formalization and modularization of compensation as a reusable service that bridges the gap between raw machine behavior and portable control logic.

%% file: 05_Conclusion.tex
\section{Discussion}
\label{sec:Discussion}
In reference to the five-dimensional digital twin model (see~Fig.~\ref{fig:AAST_T3}), 
this work extends the AAS by introducing a proactive component and a conceptual framework for service interaction.
Consequently, the establishment of standardized or shared proactive AAS components is enabled.
This is the foundation for the re-usability not only of syntax, standardized in the AAS metamodel, and semantics, standardized in the AAS submodels, but also of system behavior, potentially standardized in software containers.
Integrating AI-driven analytics and machine learning models into our proposed architecture enables the AAS to continuously analyze sensor streams, detect emerging anomalies, and autonomously trigger parameter adjustments or maintenance actions. 
This transforms traditionally reactive workflows into proactive, self-optimizing behaviors.

\section{Future Plans}
\label{sec:FuturePlans}
Realizing the case study, proof of concept~(PoC) has been provided for the implementation of proactive AAS applying a container-based approach~(cf.~TRL-3,~\cite{mankins1995technology}).
The further research roadmap is outlined from here.

\subsubsection{Full-scale Prototype}
\label{subsubsec:Prototype}
Issues such as security, responsibility and access rights have been ignored in this initial PoC.
However, for industrial or even large-scale academic application, these concerns are essential. 
Hence, a further step is to align the security concepts of the AAS with those of the container runtime.
The communication between internal services and external software components via the AAS interface is to be examined in particular. In this endeavor, a systematic service reuse across heterogeneous AASs will be crucial.
Furthermore, 
the current architecture is to be extended to include a container cluster (i.e., Kubernetes).
To this end, the service execution submodel is to be extended and the execution triggers are to be verified.

\subsubsection{Empirical Evaluation}
\label{subsubsec:EmpiricalEvaluation}
In parallel, the applicability of the concept is to be examined in a broad-based survey of AAS users.
AAS user groups and members of the IDTA, as well as publicly listed IDTA partners, will be contacted, which means approximately 150 to 180 potential survey participants.
The survey aims to identify potential use cases and collect early feedback on the container-based approach to implement proactive AAS. 

\subsubsection{Full-scale Case Study}
Finally, the applicability of the prototype is demonstrated (cf.~\ref{subsubsec:Prototype}) will be demonstrated based on a real-world industrial use case, identified through a survey (cf.~\ref{subsubsec:EmpiricalEvaluation}). 
In cooperation with the partner, the selected use case is initially set up and tested in the laboratory, then transferred to an operational environment. (cf.~TRL-4 to TRL-7,~\cite{mankins1995technology}).


\section{Conclusion}
\label{sec:Conclusion}
This paper presents the vision of reusable software services via a container-based approach in order to realize proactive AAS DTs. 
By embedding service execution logic and trigger conditions directly within the AAS structure, the proposed approach enables dynamic, context-aware behavior and paves the way for more autonomous and service-integrated digital twins.
This advancement strengthens the role of the AAS not only as a static information container but also as an active participant in the lifecycle and operation of industrial assets.
This will impact the further acceptance of AAS technology and is the first step toward the provision of standardized value-added services within the AAS ecosystem.
